\documentclass[italian,english]{article}
\usepackage[latin9]{inputenc}
\usepackage{color}
\usepackage{amsmath}
\usepackage{amssymb}

\newcommand{\lyxaddress}[1]{
\par {\raggedright #1
\vspace{1.4em}
\noindent\par}
}

\usepackage{babel}

\begin{document}

\title{\textbf{Interferometric detection of gravitational waves arising
from extended theories of gravity}}

\author{\textbf{Christian Corda }}

\maketitle

\lyxaddress{\begin{center}
Associazione Scientifica Galileo Galilei, Via Pier Cironi 16 - 59100
PRATO, Italy 
\par\end{center}}

\begin{center}
\textit{E-mail address:} \textcolor{blue}{christian.corda@ego-gw.it} 
\par\end{center}
\begin{abstract}
Even if Einstein's General Relativity achieved a great success and
overcame lots of experimental tests, it also showed some shortcomings
and flaws which today advise theorists to ask if it is the definitive
theory of gravity. In this letter Proceeding we show that, by assuming
that advanced projects on the detection of Gravitational Waves (GWs)
will improve their sensitivity, allowing to perform a GWs astronomy,
accurate angular and frequency dependent response functions of interferometers
for GWs arising from various Theories of Gravity, i.e. General Relativity
and Extended Theories of Gravity, could aim in discriminating among
various theories. 
\end{abstract}
Recently, the data analysis of interferometric GWs detectors has been
started (for the current status of GWs interferometers see \cite{key-1})
and the scientific community aims at a first direct detection of GWs
in next years. 

Detectors for GWs will be important for a better knowledge of the
Universe and also to confirm or ruling out the physical consistency
of General Relativity or of any other theory of gravitation \cite{key-2,key-3,key-4,key-5,key-6,key-7}.
This is because, in the context of Extended Theories of Gravity, some
differences between General Relativity and the others theories can
be pointed out starting by the linearized theory of gravity \cite{key-2,key-3,key-4,key-5,key-6,key-7,key-8,key-9,key-10,key-11,key-12}.
In this picture, detectors for GWs are in principle sensitive also
to a hypothetical \textit{scalar} component of gravitational radiation,
that appears in extended theories of gravity like scalar-tensor gravity
\cite{key-4,key-8,key-9,key-10,key-12}, bi-metric theory \cite{key-5},
high order theories \cite{key-2,key-3,key-6,key-7}, Brans-Dicke theory
\cite{key-13} and string theory \cite{key-14}.

Motivations of extending gravity arise from the fact that even if
Einstein's General Relativity \cite{key-15} achieved a great success
(see for example the opinion of Landau who says that General Relativity
is, together with Quantum Field Theory, the best scientific theory
of all \cite{key-16}) and overcame lots of experimental tests \cite{key-15},
it also showed some shortcomings and flaws which today advise theorists
to ask if it is the definitive theory of gravity \cite{key-17,key-18}.
Differently from other field theories like the electromagnetic theory,
General Relativity is very difficult to be quantized. This fact rules
out the possibility of treating gravitation like other quantum theories
and precludes the unification of gravity with other interactions.
At the present time, it is not possible to realize a consistent Quantum
Gravity Theory which leads to the unification of gravitation with
the other forces \cite{key-17,key-18}. 

On the other hand, one can define \textit{Extended Theories of Gravity}
those semiclassical theories where the Lagrangian is modified, in
respect to the standard Einstein-Hilbert gravitational Lagrangian,
adding high-order terms in the curvature invariants (terms like $R^{2}$,
$R^{\alpha\beta}R_{\alpha\beta}$, $R^{\alpha\beta\gamma\delta}R_{\alpha\beta\gamma\delta}$,
$R\Box R$, $R\Box^{k}R$) or terms with scalar fields non minimally
coupled to geometry (terms like $\phi^{2}R$) \cite{key-17,key-18}.
In general, one has to emphasize that terms like those are present
in all the approaches to perform the unification between gravity and
other interactions. More, from a cosmological point of view, such
modifies of General Relativity generate inflationary frameworks which
are very important as they solve lots of problems of the Standard
Universe Model \cite{key-19}. Note that we are not telling that General
Relativity is wrong. It is well known that, even in the context of
Extended Theories, General Relativity remains the most important part
of the structure \cite{key-4,key-7,key-17,key-18}. We are only trying
to understand if weak modifies to such a structure could be needed
to solve some theoretical and observable problems \cite{key-17,key-18}.
In this picture, we also recall that even Einstein told that General
Relativity could not be definitive \cite{key-28}. In fact, during
his famous research on the Unified Field Theory, he tried to realize
a theory that he called {}``Generalized Theory of Gravitation'',
and he said that mathematical difficulties precluded him to obtain
the final equations \cite{key-28}.

In the general context of cosmological evidences, there are also other
considerations, which suggest an extension of General Relativity.
As a matter of fact, the accelerated expansion of the Universe, which
is today observed, shows that cosmological dynamic is dominated by
the so called Dark Energy, which gives a large negative pressure.
This is the standard picture, in which such new ingredient is considered
as a source of the \textit{right side} of the field equations. It
should be some form of non-clustered and non-zero vacuum energy which,
together with the clustered Dark Matter, drives the global dynamics.
This is the so called {}``concordance model'' ($\Lambda$CDM) which
gives, in agreement with the CMBR, LSS and SNeIa data, a good tapestry
of the today observed Universe, but presents several shortcomings
as the well known {}``coincidence'' and {}``cosmological constant''
problems \cite{key-20}. An alternative approach is changing the \textit{left
side} of the field equations, seeing if observed cosmic dynamics can
be achieved extending General Relativity \cite{key-17,key-18,key-21,key-22,key-23}.
In this different context, it is not required to find out candidates
for Dark Energy and Dark Matter, that, till now, have not been found,
but only the {}``observed'' ingredients, which are curvature and
baryon matter, have to be taken into account. Considering this point
of view, one can think that gravity is different at various scales
\cite{key-21} and a room for alternative theories is present. In
principle, the most popular Dark Energy and Dark Matter models can
be achieved considering $f(R)$ theories of gravity, where $R$ is
the Ricci curvature scalar, and/or Scalar-Tensor Gravity \cite{key-17,key-18,key-22,key-23}.

In this letter Proceeding we show that, by assuming that advanced
projects on the detection of GWs will improve their sensitivity allowing
to perform a GWs astronomy \cite{key-1}, accurate angular and frequency
dependent response functions of interferometers for gravitational
waves arising from various Theories of Gravity, i.e. General Relativity
and Extended Theories of Gravity, could aim in discriminating among
various theories \cite{key-2,key-3,key-4,key-5,key-6,key-7,key-10,key-11,key-12}. 

The line element for GWs arising from standard General Relativity
is given by \cite{key-1,key-15,key-24,key-25} (note that in this
letter Proceeding we work with $G=1$, $c=1$ and $\hbar=1$)

\begin{equation}
ds^{2}=dt^{2}-dz^{2}-(1+h_{+})dx^{2}-(1-h_{+})dy^{2}-2h_{\times}dxdy,\label{eq: metrica TT totale}\end{equation}

where $h_{+}(t+z)$ and $h_{\times}(t+z)$ are the weak perturbations
due to the $+$ and the $\times$ polarizations which are expressed
in terms of synchronous coordinates in the Transverse Traceless (TT)
gauge. Using the {}``bouncing photon'' analysis, in \cite{key-24,key-25}
it has been shown that the total frequency and angular dependent response
function (i.e. the detector pattern) of an interferometer to the $+$
polarization of the GW is:\begin{align}
\tilde{H}^{+}(\omega) & \equiv\Upsilon_{u}^{+}(\omega)-\Upsilon_{v}^{+}(\omega)\nonumber \\
 & =\frac{(\cos^{2}\theta\cos^{2}\phi-\sin^{2}\phi)}{2L}\tilde{H}_{u}(\omega,\theta,\phi)-\frac{(\cos^{2}\theta\sin^{2}\phi-\cos^{2}\phi)}{2L}\tilde{H}_{v}(\omega,\theta,\phi),\label{eq: risposta totale Virgo +}\end{align}

that, in the low frequencies limit ($\omega\rightarrow0$) gives the
well known low frequency response function of \cite{key-26,key-27}
for the $+$ polarization: 

\begin{equation}
\tilde{H}^{+}(\omega)=\frac{1}{2}(1+\cos^{2}\theta)\cos2\phi+O\left(\omega\right)\,.\label{eq: risposta totale approssimata}\end{equation}

The same analysis works for the $\times$ polarization (see \cite{key-24,key-25}
for details). One obtains that the total frequency and angular dependent
response function of an interferometer to the $\times$ polarization
is:

\begin{equation}
\tilde{H}^{\times}(\omega)=\frac{-\cos\theta\cos\phi\sin\phi}{L}[\tilde{H}_{u}(\omega,\theta,\phi)+\tilde{H}_{v}(\omega,\theta,\phi)],\label{eq: risposta totale Virgo per}\end{equation}
that, in the low frequencies limit ($\omega\rightarrow0$), gives
the low frequency response function of \cite{key-26,key-27} for the
$\times$ polarization: \begin{equation}
\tilde{H}^{\times}(\omega)=-\cos\theta\tilde{H}_{v}(\omega,\theta,\phi)\sin2\phi+O\left(\omega\right)\,.\label{eq: risposta totale approssimata 2}\end{equation}

In these equations $u$ and $v$ are the directions of the interferometer
arms in respect to the propagating GW and we have defined \cite{key-24,key-25}

\begin{equation}
\tilde{H}_{u}(\omega,\theta,\phi)\equiv\frac{-1+\exp(2i\omega L)-\sin\theta\cos\phi((1+\exp(2i\omega L)-2\exp i\omega L(1-\sin\theta\cos\phi)))}{2i\omega(1+\sin^{2}\theta\cos^{2}\phi)}\label{eq: fefinizione Hu}\end{equation}

and 

\begin{align}
\tilde{H}_{v}(\omega,\theta,\phi) & \equiv\frac{-1+\exp(2i\omega L)-\sin\theta\sin\phi((1+\exp(2i\omega L)-2\exp i\omega L(1-\sin\theta\sin\phi)))}{2i\omega(1+\sin^{2}\theta\sin^{2}\phi)}.\label{eq: fefinizione Hv}\end{align}

The case of massless Scalar-Tensor Gravity has been discussed in \cite{key-4,key-12}.
In this case, the line-element in the TT gauge can be extended with
one more polarization, labelled with $\Phi(t+z)$, i.e.

\begin{equation}
ds^{2}=dt^{2}-dz^{2}-(1+h_{+}+\Phi)dx^{2}-(1-h_{+}+\Phi)dy^{2}-2h_{\times}dxdy.\label{eq: metrica TT super totale}\end{equation}

Now, the total frequency and angular dependent response function of
an interferometer to this {}``scalar'' polarization is \cite{key-4,key-12}\begin{align}
\tilde{H}^{\Phi}(\omega) & =\frac{\sin\theta}{2i\omega L}\{\cos\phi[1+\exp(2i\omega L)-2\exp i\omega L(1+\sin\theta\cos\phi)]+\nonumber \\
 & -\sin\phi[1+\exp(2i\omega L)-2\exp i\omega L(1+\sin\theta\sin\phi)]\}\,,\label{eq: risposta totale Virgo scalar}\end{align}

that, in the low frequencies limit ($\omega\rightarrow0$), gives
the low frequency response function of \cite{key-9,key-14} for the
$\Phi$ polarization: \textbf{\begin{equation}
\tilde{H}^{\Phi}(\omega)=-\sin^{2}\theta\cos2\phi+O(\omega).\label{eq: risposta totale approssimata scalar}\end{equation}
}

In \cite{key-2,key-3,key-4,key-7} it has also been shown that, in
the framework of GWs, the cases of massive Scalar-Tensor Gravity and
$f(R)$ theories are totally equivalent (this is not surprising as
it is well known that there is a more general conformal equivalence
between Scalar-Tensor Gravity and $f(R)$ theories, even if there
is a large debate on the possibility that such a conformal equivalence
should be a \emph{physical} equivalence too \cite{key-17,key-18,key-21}).
In such cases, because of the presence of a small mass, a longitudinal
component is present in the third polarization, thus it is impossible
to extend the TT gauge to the third mode \cite{key-2,key-3,key-4,key-6,key-7}.
But, by using gauge transformations, one can put the line-element
due to such a third scalar mode in a conformally flat form \cite{key-2,key-3,key-4,key-6,key-7}: 

\begin{equation}
ds^{2}=[1+\Phi(t,z)](-dt^{2}+dz^{2}+dx^{2}+dy^{2}).\label{eq: metrica puramente scalare}\end{equation}

If the interferometer arm is parallel to the propagating GW the longitudinal
response function associated to such a massive mode is \cite{key-2,key-7}
\begin{equation}
\begin{array}{c}
\Upsilon_{l}(\omega)=\frac{1}{m^{4}\omega^{2}L}(\frac{1}{2}(1+\exp[2i\omega L])m^{2}\omega^{2}L(m^{2}-2\omega^{2})+\\
\\-i\exp[2i\omega L]\omega^{2}\sqrt{-m^{2}+\omega^{2}}(4\omega^{2}+m^{2}(-1-iL\omega))+\\
\\+\omega^{2}\sqrt{-m^{2}+\omega^{2}}(-4i\omega^{2}+m^{2}(i+\omega L))+\\
\\+\exp[iL(\omega+\sqrt{-m^{2}+\omega^{2}})](m^{6}L+m^{4}\omega^{2}L+8i\omega^{4}\sqrt{-m^{2}+\omega^{2}}+\\
\\+m^{2}(-2L\omega^{4}-2i\omega^{2}\sqrt{-m^{2}+\omega^{2}}))+2\exp[i\omega L]\omega^{3}(-3m^{2}+4\omega^{2})\sin[\omega L]),\end{array}\label{eq: risposta totale lungo z massa}\end{equation}

where $m$ is the small mass associated to the GW.

Thus, by assuming that advanced projects on the detection of GWs will
improve their sensitivity allowing to perform a GWs astronomy (this
is due because signals from GWs are quite weak) \cite{key-1}, one
will only have to look the interferometer response functions to understand
which is the correct theory of gravity. If only the two response functions
(\ref{eq: risposta totale Virgo +}) and (\ref{eq: risposta totale Virgo per})
will be present we will conclude that General Relativity is the definitive
theory of gravitation. If the response function (\ref{eq: risposta totale Virgo scalar})
will be present too, we will conclude that massless Scalar - Tensor
Gravity is the correct theory of gravitation. Finally, if the third
response function will be given by Eq. (\ref{eq: risposta totale lungo z massa}),
we will learn that the correct theory of gravity will be Scalar -
Tensor Gravity which is equivalent to $f(R)$ theories. In any case,
such response functions will help in discriminating between gravity
theories. This is because General Relativity is the only gravity theory
which admits only the two response functions (\ref{eq: risposta totale Virgo +})
and (\ref{eq: risposta totale Virgo per}) \cite{key-4,key-7,key-17,key-18}.
Such response functions correspond to the two {}``canonical'' polarizations
$h_{+}$ and $h_{\times}.$ Thus, if a third polarization will be
present, a third response function will be detected by GWs interferometers
and this fact will rule out General Relativity like the definitive
theory of gravity confirming Einstein's intuition that a modify is
needed \cite{key-28}.

\subsubsection*{Conclusion remarks}

In this review letter Proceeding we have shown that, by assuming that
advanced projects on the detection of GWs will improve their sensitivity,
allowing to perform a GWs astronomy, accurate angular and frequency
dependent response functions of interferometers for gravitational
waves arising from various Theories of Gravity, i.e. General Relativity
and Extended Theories of Gravity, will be an important help in discriminating
among various gravity theories.

\subsubsection*{Acknowledgments}

I strongly thank Professor Remo Ruffini for his kind invitation to
the 3rd Stueckelberg Workshop.


\begin{thebibliography}{28}
\bibitem{key-1}A. Giazotto - Journ. of Phys., Conf. Series 120, 032002
(2008) 

\bibitem[2]{key-2}C. Corda - J. Cosmol. Astropart. Phys. JCAP04009
(2007)

\bibitem[3]{key-3}\foreignlanguage{italian}{C. Corda - Int. Journ.}
Mod. Phys. A 23, 10, 1521-1535 (2008)

\bibitem[4]{key-4}Capozziello S and C. Corda - Int. J. Mod. Phys.
D \textbf{15,} 1119 -1150 (2006) 

\bibitem[5]{key-5}C. Corda - Astropart. Phys. 28, 2, 247-250 (2007)

\bibitem[6]{key-6}C. Corda - Astropart. Phys. 30, 4 209-215 (2008)

\bibitem[7]{key-7}S. Capozziello, C. Corda and M. F. De Laurentis
- Phys.Lett. B, 669, 5, 255-259, (2008) 

\bibitem[8]{key-8}T. Damour and G. Esposito-Farese - Class. Quant.
Grav. \textbf{9} 2093-2176 (1992);  

\bibitem[9]{key-9}M. E. Tobar , T. Suzuki and K. Kuroda \foreignlanguage{italian}{Phys.}
Rev. \foreignlanguage{italian}{D 59 102002 (1999)}

\bibitem[10]{key-10}S. Capozziello, C. Corda and M. F. De Laurentis
- Mod. Phys. Lett. A 22, 35, 2647-2655 (2007)

\bibitem[11]{key-11}S. Capozziello, C. Corda and M. F. De Laurentis
- Mod. Phys. Lett. A 22, 15, 1097-1104 (2007)

\bibitem[12]{key-12}C. Corda - Mod. Phys. Lett. A No. 22, 23, 1727-1735
(2007)

\bibitem[13]{key-13}C. Brans and R. H. Dicke - Phys. Rev. 124, 925
(1961)

\bibitem[14]{key-14}N. Bonasia and M. Gasperini - Phys. Rew. D \textbf{71}
104020 (2005)

\bibitem[15]{key-15}\foreignlanguage{italian}{C. W. Misner, K. S.
Thorne and J. A. Wheeler - {}``Gravitation'' - W.H.Feeman and Company
- 1973} 

\bibitem[16]{key-16}L. Landau and E. Lifsits - {}``Teoria dei campi''
- Editori riuniti edition III (1999) 

\bibitem[17]{key-17}S. Capozziello and M. Francaviglia - Gen. Rel.
Grav. 40, 2-3, 357 (2008)

\bibitem[18]{key-18}V. Faraoni and T. P. Sotiriou - arXiv:0805.1249,
to appear in Rev. Mod. Phys.

\bibitem[19]{key-19}A. Starobinsky - Phys. Lett. B, 91, e 1, 99-102
(1980)

\selectlanguage{italian}%
\bibitem[20]{key-20}\foreignlanguage{english}{P. J. E. Peebles and
B. Ratra - Rev. Mod. Phys. 75 8559 (2003) }

\selectlanguage{english}%
\bibitem[21]{key-21}G. Cognola, E. Elizalde, S. Nojiri, S.D. Odintsov,
L. Sebastiani, S. Zerbini - Phys. Rev. D 77, 046009 (2008) 

\bibitem[22]{key-22}S. Nojiri and S.D. Odintsov - Int. J. Geom. Meth.Mod.Phys.4:115-146
(2007) 

\bibitem[23]{key-23}E. Elizalde, P. J. Silva - Phys. Rev. D78, 061501
(2008)

\bibitem[24]{key-24}C. Corda - Astropart. Phys. \textbf{27,} No 6,
539-549 (2007)

\bibitem[25]{key-25}C. Corda - Int. J. Mod. Phys. D \textbf{16,}
9, 1497-1517  (2007)

\bibitem[26]{key-26}K. S. Thorne - \textit{300 Years of Gravitation}
- Ed. Hawking SW and Israel W Cambridge University Press p. 330 (1987)

\bibitem[27]{key-27}P. Saulson - \textit{Fundamental of Interferometric
Gravitational Waves Detectors} - World Scientific, Singapore (1994) 

\bibitem[28]{key-28} A. Pais - \textit{Subtle is the Lord, The Science
and the Life of Albert Einstein} - Oxford University Press (2005)
\end{thebibliography}
\end{document}